\documentclass[12pt]{article}

\usepackage{graphics, color}
\usepackage{graphicx}
\usepackage{amssymb}
\pdfoutput=1

\newcommand{\sect}[1]{\section{#1}\setcounter{equation}{0}}

\def\gsim{\, \rlap{$>$}{\lower 1.1ex\hbox{$\sim$}}\,}
\def\lsim{\, \rlap{$<$}{\lower 1.1ex\hbox{$\sim$}}\,}

\def\w{\omega}

 \newcommand{\be}{\begin{equation}}
\newcommand{\ee}{\end{equation}}
 \newcommand{\bal}{\begin{align}}
 \newcommand{\eal}{\end{align}}
\newcommand{\ben}{\begin{equation*}}
\newcommand{\een}{\end{equation*}}
\newcommand{\bea}{\begin{eqnarray}}
\newcommand{\eea}{\end{eqnarray}}
\newcommand{\bean}{\begin{eqnarray*}}
\newcommand{\eean}{\end{eqnarray*}}
\newcommand{\bes}{\begin{subequations}}
\newcommand{\ees}{\end{subequations}}
\newcommand{\ra}{\rightarrow}

\usepackage{epsfig}

\newcommand{\comment}[1]{}

\begin{document}

\begin{titlepage}
\bigskip
\rightline{}

\bigskip
\bigskip\bigskip\bigskip
\centerline{\Large \bf Surprising Connections Between}
\bigskip
\centerline{\Large \bf General Relativity and Condensed Matter}

\bigskip\bigskip\bigskip
\bigskip\bigskip\bigskip

\centerline{{\large Gary T. Horowitz}}
\medskip
\centerline{\em Department of Physics}
\centerline{\em University of California}
\centerline{\em Santa Barbara, CA 93106-4030}\bigskip

\bigskip
\bigskip\bigskip


\begin{abstract}
This brief review is intended to introduce gravitational physicists to recent developments in which general relativity is being  used to describe certain aspects of condensed matter systems, e.g., superconductivity.

\end{abstract}
\end{titlepage}
\baselineskip = 17pt
\setcounter{footnote}{0}
\setcounter{equation}{0}
\sect{Introduction}

Over the past few years a surprising connection has been found between general relativity and nongravitational physics, including condensed matter. In fact, we will present evidence for the following remarkable claim:

{\it In addition to describing gravitational phenomena (black holes, gravitational waves, etc.) general relativity can also describe other fields of physics, including aspects of superconductivity. }

 We begin by  reviewing black hole thermodynamics since that will play an important role in our discussion. In the next section, 
we  describe the remarkable gauge/gravity duality which is the main tool that will allow us to use general relativity to describe other areas of physics.
Finally, in section three we present the example of  superconductivity.
  
  Following the seminal work by Bekenstein and Hawking, it is known that black holes are thermodynamic objects with a temperature  related to the surface gravity $\kappa$ by $T = \hbar \kappa/2\pi$,  and entropy $S = A/4\hbar G $ where $A$ is the area of the event horizon.   Although black hole thermodynamics is often studied with asymptotically flat boundary conditions, this is really not suitable for discussing equilibrium configurations. 
If you try to put a black hole in thermal equilibrium with its Hawking radiation, there is nonzero energy density out to infinity so the solution is not really asymptotically flat. Instead it resembles a cosmology and starts to expand or contract. 
  
The situation is much better  with anti de Sitter (AdS) boundary conditions. AdS is the maximally symmetric solution of Einstein's equation with negative cosmological constant. 
It has constant negative curvature and this negative curvature acts like a confining box, so there are static black holes in thermal equilibrium with their Hawking radiation \cite{Hawking:1982dh}.
 The study of black holes  is also much richer in AdS. Unlike asymptotically flat black holes which must be spherical, AdS black holes  exist with   spherical, planar, or hyperbolic  horizon geometry.
  
  Spherical black holes have a minimum temperature in AdS set by the cosmological constant: $T_{min} \sim 1/L$ where $\Lambda = - 3/L^2$. ($L$ is called the radius of curvature of AdS.) Small black holes act just like the asymptotically flat Schwarzschild solution. Their temperature is  $T \sim1/r_0$ where $r_0$ is the horizon radius, and they can evaporate. However large black holes are different and have a temperature which grows linearly with $r_0$. This means that large black holes have positive specific heat: Their temperature increases when you add mass. So the  equilibrium configurations of black holes and radiation are stable.  For $T<1/L$ the equilibrium configuration is just a gas a particles in AdS. There is a (first order) phase transition between this gas of particles  at low $T$ and a large black holes at high $T$ which is called the Hawking-Page transition \cite{Hawking:1982dh}. 
  
   We will be more interested in the planar black hole. It has a metric which is static and translationally invariant. Setting $L = 1$ for convenience, it is  given by
   \be\label{metric}
   ds^2 = r^2[-f(r) dt^2 + dx_idx^i] + {dr^2 \over r^2 f(r) }
   \ee
 where $i = 1,2$ and
 \be f(r) = 1-{r_+^3\over r^3}
 \ee
   This is a solution to Einstein's equation with negative cosmological constant. The horizon is at $r=r_+$ and $r_+ =0 $ is AdS itself. Dropping numerical factors,
   the Hawking temperature is $T \sim  r_+$ (for all $r_+$).
The total energy is $E \sim r_+^3 V \sim T^3 V$
and the entropy is $S \sim A \sim r_+^2 V \sim T^2 V$. 
So not only is the specific heat positive, the energy and entropy are exactly like a thermal gas in 2+1 dimensions!  This is the first indication that general relativity with AdS boundary conditions can be related to nongravitational systems.

Charged (planar) black holes in AdS are similar to Reissner-Nordstrom. The metric is of the same form as (\ref{metric}) with 
 \be f(r) = 1-(1 + b){r_+^3\over r^3} + b{r_+^4\over r^4}
 \ee
 The vector potential is
 \be A_t = \mu - {\rho\over r}
 \ee
 where $\rho$ is the charge density and $\mu$ is chosen so that $A_t$ vanishes at the horizon. (This is required in order for $A_\nu A^\nu$ to remain finite there.) The parameters $b,r_+$ in the metric are related to $\mu$ and $\rho$, so the solutions are labeled by only  two parameters, which can be thought of as the mass and charge.
There is an extremal limit corresponding to the maximum charge for a given mass with $T = 0$ and nonzero entropy.
Having $S \ne 0$ at $T = 0$ is uncommon for a thermodynamic system and seems to imply a highly degenerate ground state. We will see that the Reissner-Nordstrom AdS solution is often unstable at low temperature when coupled to other matter (e.g. a charged scalar field) and the new black hole has $S = 0$ at $T = 0$.

 \setcounter{equation}{0}  
 \section{Gauge/gravity duality}

   We begin with a few comments about gauge theory and string theory. Gauge theories  are generalizations of electromagnetism in which the familiar U(1) gauge symmetry is replaced by a noncommuting group such as SU(N). Our standard model of particle physics is based on a gauge theory. In particular, QCD (the theory of the strong interactions) has SU(3) gauge symmetry. The interactions are weak at high energy but become strong at low energy causing quark confinement.
   
      String theory is based on the idea that elementary particles are not pointlike, but excitations of a one dimensional string.  Strings interact with a simple splitting and joining interaction. It has been shown that string theory includes gravity and in fact reduces to general relativity (with certain matter) in a classical limit.
   
   `t Hooft argued in the 1970's that a 1/N expansion of an SU(N) gauge theory would resemble a theory of strings \cite{'tHooft:1973jz}. This was based on the structure of the perturbation theory. This turns out to be correct, but
 it took more than twenty years for this idea to be made precise.
 The precise formulation takes the form of a duality \cite{Maldacena:1997re}:
  
  {\it Gauge/gravity duality: With anti de Sitter boundary conditions, string theory (which includes gravity) is completely equivalent to a (nongravitational) gauge theory living on the boundary at infinity. }
   
   Of course these theories look very different at weak coupling, but the point is that
when the string theory is weakly coupled, the gauge theory is strongly coupled, and vice versa.
   This duality is a manifestation of the idea that quantum gravity (at least with AdS boundary conditions) is holographic: It can be completely described by degrees of freedom living in a lower dimensional space. This idea was first discussed by `t Hooft \cite{'tHooft:1993gx} and Susskind \cite{Susskind:1994vu}.
Even though this duality has not been rigorously proved, there is by now very strong  evidence that it is true. 
   
  Setting $r_+ = 0$ in (\ref{metric}) we can write anti de Sitter space in planar coordinates   \be\label{AdS}
    ds^2 = r^2[-dt^2 + dx_idx^i] + {dr^2 \over r^2  } 
   \ee
Conformally rescaling this metric by $1/r^2$, one can smoothly extend the geometry to the boundary at $r = \infty$. The metric on the boundary is just Minkowski spacetime, and the gauge theory lives on this spacetime. (The physical spacetime away from the boundary is often refered to as the ``bulk".) The metric (\ref{AdS}) is invariant under a scaling symmetry:
\be\label{scaling}
 r \rightarrow      ar,\quad  (t, x^i) \ra     (t/a, x^i/a) 
\ee
It follows that small $r$ corresponds to large distances or low energy in the gauge theory.    
   
One of the main applications of gauge/gravity duality  has been to gain new insight into strongly coupled gauge theories. In particular, it provides a
simple geometric picture of confinement, which can be seen as follows.
 The potential between two quarks separated by $\Delta x^i$ is obtained from the length of a string in the bulk with endpoints at the two quarks on the boundary\footnote{Since the boundary is at infinity, the actual length is infinite, but this just corresponds to an infinite self energy of the pointlike quarks. There is a well defined way to subtract this divergence and the resulting finite quantity defines the potential.}. Since the string is under tension, it wants to minimize its length. 
Since the geometry is warped, as $\Delta x^i$ increases, the minimal curve extends farther in to smaller radii. In fact, in AdS itself, 
 the length of this curve is independent of $\Delta x^i$ due to the scaling symmetry (\ref{scaling}). The potential vanishes and AdS describes the vacuum of a scale invariant theory.
   
   The gravity dual of a confining vacuum differs from AdS in that the spacetime ends at nonzero radius. (Recall that small radius describes low energy.)
This can happen, e.g., if a compact extra dimension smoothly caps off.
In this case, when $\Delta x^i$ becomes large, the string drops down to the minimum radius keeping $x^i$ nearly constant. It then moves a distance $\Delta x^i$ and returns to the boundary again along a curve of nearly constant $x^i$. So the length of the string grows linearly with $\Delta x^i$ when it becomes large. This gives a linearly growing potential between quarks, i.e., confinement! In this way one obtains a 
simple geometric description of a complicated quantum field theory phenomenon.

It is easy to construct a solution describing a confining vacuum. We start with the five dimensional analog of the planar AdS black hole. The metric is again given by (\ref{metric}) with now $i=1,2,3$ and
 \be f(r) = 1-{r_+^4\over r^4}
 \ee
 We now analytically continue $t = i\chi$. As usual, when constructing a euclidean black hole metric, the space only exists for $r\ge r_+$ and $\chi$ must be periodic with period $\beta = 1/T$ in order to avoid a conical singularity at $r=r_+$.  We now analytically continue $x_3 = it$ to obtain a new Lorentzian solution. The result is
  \be\label{adssoliton}
   ds^2 = r^2[f(r) d\chi^2 + dx_idx^i-dt^2] + {dr^2 \over r^2 f(r) }
   \ee
with $i=1,2$.   This metric is globally static and nonsingular. It is called the AdS soliton \cite{Horowitz:1998ha,Witten:1998zw}. The dual theory is a four dimensional gauge theory compactified on a circle (parameterized by $\chi$). At low energies it reduces to a  $2+1$ gauge theory which has a confining vacuum. The AdS soliton is the dual gravitational description of this state.

One can also turn the duality around and use the gauge theory to gain insight into
quantum gravity. In particular, it provides answers to  fundamental questions about  quantum properties of black holes. For example, it has been shown that the gauge theory has enough microstates to reproduce the Bekenstein-Hawking entropy of black holes in AdS. This explains the origin of the enormous entropy arising in black hole thermodynamics. Furthermore, since the entire process of forming a small black hole in AdS and letting it evaporate can be represented by standard Hamiltonian evolution in the gauge theory, it is clear that there is no loss of information. Black hole evaporation does not violate quantum mechanics.
After arguing against this for thirty years, Hawking finally conceded this point in 2005 \cite{Hawking:2005kf}. However, the actual mechanism by which the information gets out of the black hole is still under active investigation.
   
   So far our discussion has included quantum gravity and string theory. However, in a certain (large $N$) limit, all stringy and quantum effects are suppressed and the bulk  theory is just general relativity
(with AdS boundary conditions and possibly higher dimensions).
If gauge/gravity duality is correct, general relativity must be able to  reproduce other areas of physics. 

 One check comes from considering the   renormalization group (RG). The RG flow in a quantum field theory corresponds to obtaining an effective low energy action by integrating out high energy modes.
This corresponds to radial dependence on the gravity side. One can now compare two separate calculations. On the gauge theory side, 
one adds a mass term to the action and follows the RG flow to low energies to obtain a new field theory. On the gravity side, adding a mass term to the dual field theory correspond to modifying the
 boundary conditions for certain matter fields coupled to gravity. So one  looks for static solutions to Einstein's equation with these modified boundary conditions.
Remarkably, one finds detailed numerical agreement between the small $r$ behavior of the gravity solution and the endpoint of the RG flow \cite{Freedman:1999gp,de Boer:1999xf}.
   
   Recently, people have started to extend the rules of gauge/gravity duality to other nongravitational systems besides gauge theories.
The correspondence is not as well established in this case, but there is growing evidence that it still works. 
Starting in 2007, it was shown that there were applications of gauge/gravity duality to condensed matter physics. The Hall effect and Nerst effect were reproduced using general relativity \cite{Hartnoll:2007ai,Hartnoll:2007ih}.  (The Nerst effect is the existence of a transverse voltage arising from an applied temperature gradient in the presence of a magnetic field.) The basic idea is that a nongravitational system in thermal equilibrium at temperature $T$ is described by a black hole in AdS with Hawking temperature $T$. Using linear response theory, the electrical and thermal transport properties in the nongravitational system are obtained from linear perturbations of the black hole.
Given this early success, it is natural to ask: Is there a gravitational description of superconductivity?

\setcounter{equation}{0}   
   \section{Superconductivity}
   
    In conventional superconductors (e.g., Al, Nb, Pb) pairs of elections with opposite spin can bind to form a charged boson called a Cooper pair.
 Below a critical temperature $T_c$, there is a second order phase transition and these bosons condense.
 The DC conductivity becomes infinite resulting in superconductivity.
 This is well described by BCS theory \cite{Bardeen:1957mv} in which the binding is due to interactions with lattice vibrations or phonons.  This interaction is weak and a typical Cooper pair is much larger than the lattice spacing. 
Particle-like excitations above the superconducting ground state are usually called quasiparticles.
    
A new class of high $T_c$ superconductors were discovered in 1986 \cite{Bednorz:1986tc}. They are cuprates and the superconductivity is along the $CuO_2$ planes. 
 The highest $T_c$ known today (at atmospheric pressure)  is $T_c = 134^oK$ for a mercury, barium, copper oxide compound.  If you apply pressure, $T_c$ climbs to about 160K.
Another class of superconductors were discovered in 2008 based on iron and not copper \cite{Hosono:2008}. The  highest $T_c$ so far is $56^oK$. These materials are also layered and the superconductivity is again associated with the two dimensional planes.  They are called  iron pnictides since they involve other elements like arsenic  in the nitrogen group of the periodic table.
There is evidence that electron pairs still form in these high $T_c$ materials, but the pairing mechanism is not well understood. Unlike BCS theory, the coupling is not weak.

One might hope that gauge/gravity duality can be used to gain some insight into these high temperature superconductors. The first step is to find a gravitational dual of a  superconductor. This was found in \cite{Hartnoll:2008vx} (for more complete reviews, see \cite{Horowitz}).
 The minimal ingredients are the following.
      In a superconductor we need the notion of temperature and condensate. On the gravity side the role of temperature is played by a black hole with the Hawking temperature of the black hole equal to the field theory temperature. The role of a charged condensate will be played by a charged scalar field. So we need to find a black hole that has scalar hair at low temperatures but no hair at high temperatures. At first sight this does not sound like an easy task. 
There are no hair theorems which say that certain matter fields must be trivial outside a black hole. The field usually wants to fall into the black hole or radiate out to infinity. But these theorems usually apply to linear fields coupled to gravity. 
It is known that you can have hair for certain nonlinear theories. 

A surprisingly simple solution to this problem was found by Gubser \cite{Gubser:2008px}. He argued that a charged scalar field around a {\it charged} black hole in AdS would have the desired property. Consider
\be\label{action}
 S=\int d^{4}x \sqrt{-g}\left(R + {6} -\frac{1}{4}F_{\mu \nu}F^{\mu \nu} - |\nabla\Psi-i qA\Psi|^2 - m^2|\Psi^2|\right). 
\ee
This is just general relativity with a negative cosmological constant $\Lambda = -3$ (again setting the AdS radius to one), coupled to a Maxwell field and charged scalar with mass $m$ and charge $q$. It is easy to see why black holes in this theory might be unstable to forming scalar hair:
For an electrically charged black hole, the effective mass of $\Psi$ is
\be
 m^2_{eff} = m^2 + q^2 g^{tt} A_t^2.
 \ee
  The last term is negative, so there is a chance that $m^2_{eff}$ becomes sufficiently negative near the horizon to destabilize the scalar field.

We will see that black holes in this theory indeed develop scalar hair at low temperature. These four dimensional hairy black holes are dual to $2+1$ dimensional field theories so it is the right context to try to understand the superconductivity along the two dimensional planes. One might wonder why such a simple type of hair was not noticed earlier. A possible explanation is that this does not work for asymptotically flat black holes. The AdS boundary conditions are crucial. 
 One way to understand the difference is by the following quantum argument. Let $Q_i$ be the initial charge on the black hole. 
 If $qQ_i$ is large enough, even maximally charged black holes with zero Hawking temperature create pairs of charged particles. This is simply due to the fact that the electric field near the horizon is strong enough to pull pairs of oppositely charged particles out of the vacuum via the Schwinger mechanism of ordinary field theory.  The particle with opposite charge to the black hole falls into the horizon, reducing $Q_i$ while the particle with the same sign charge as  the black hole is repelled away. In asymptotically flat spacetime, these particles escape to infinity, so the final result is a standard Reissner-Nordstrom black hole with final charge $Q_f < Q_i$. In AdS, the charged particles cannot escape since the negative cosmological constant acts like a confining box, and they settle outside the horizon. This gas of charged particles is the quantum description of the hair. This quantum process has an entirely classical analog in terms of superradiance of the charged scalar field.

To find the hairy black holes, we start with  the following ansatz for the metric
\be\label{eq:metric}
 ds^2=-g(r) e^{-\chi(r)} dt^2+{dr^2\over g(r)}+r^2(dx^2+dy^2)
\ee
and the following ansatz for the matter fields:
\be
A=\phi(r)~dt, \quad \Psi = \psi(r)
\ee
Substituting into the field equations yield four coupled nonlinear ordinary differential equations which can be solved numerically (for some analytic results, see \cite{Siopsis:2010uq,Gregory:2009fj}). At the horizon, $r = r_+$, $g$ and $\phi$ vanish, and $\chi$ and $\psi$ are constant. Asymptotically, the metric approaches (\ref{AdS}) and the Maxwell potential approaches
\be\label{asympphi}
\phi = \mu - \frac{\rho}{r} + \cdots \,.
\ee
The asymptotic behavior of the scalar field depends on the mass. For $m^2 = -2$ we have
      \be\label{asymppsi}
\psi =  \frac{\psi^{(2)}}{r^2} + \cdots \,.
\ee
Gauge/gravity duality relates this asymptotic behavior to properties of the dual field theory:
$\mu$ is the chemical potential and $\rho$  is the charge density.
 There is a dimension two operator dual to $\psi$, and its expectation value is $O_2 =  \psi^{(2)}$.
 
 We want to know how the condensate $ O_2$ behaves as a function of temperature. But there is a scaling symmetry analogous to (\ref{scaling}) which applies to the hairy black holes. This will trivially rescale the temperature and condensate. It is convenient to use the chemical potential to fix a  scale and consider $\sqrt { O_2}/\mu$ as a function of $T/\mu$. When one does this, one finds that the condensate is nonzero only when $T/\mu$ is small enough.  Setting $T_c$ equal to the critical temperature when the condensate first turns on,  we get Fig. 1. This curve is  qualitatively similar to that
obtained in BCS theory, and observed in many materials, where the
condensate rises quickly and goes to a constant at zero temperature.

The critical temperature is proportional to $\mu$, which is not surprising since it is the only other scale in the problem.
      Near $T_c$, there is a square root behavior $ O_2 \sim(1 - T/T_c)^{1/2}$. This is  standard mean field behavior predicted by Landau-Ginzburg theory.  As $T\ra   0$, one finds that the area of the horizon of the hairy black hole vanishes, consistent with a unique ground state.

   \begin{figure}\begin{center}
\includegraphics[width=.7\textwidth]{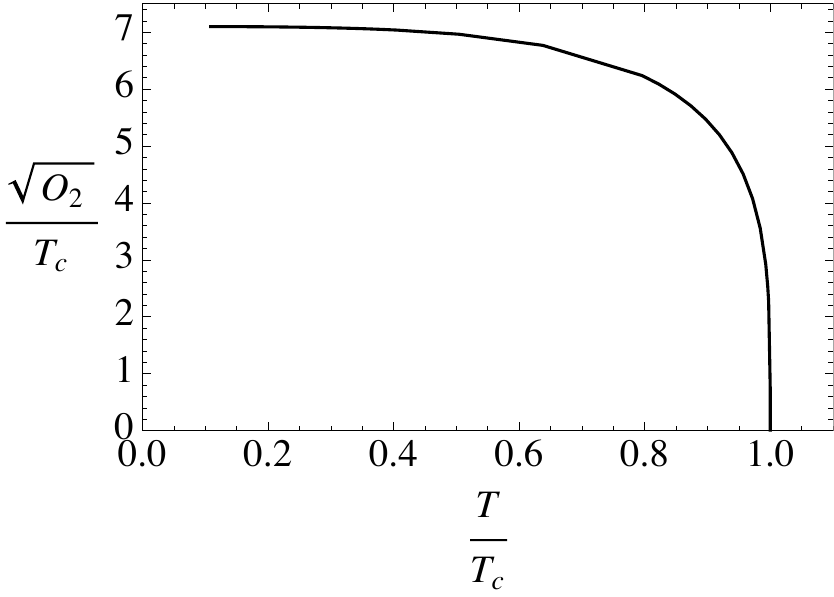}
\caption{The condensate as a function of temperature. The critical temperature is proportional to the chemical potential.}\label{cond2}
\end{center}\end{figure}

 One can compute the free energy (euclidean action) of these hairy configurations and compare with the solution without hair: $\psi = 0$,  $A_t = \mu - \rho/r$. The free energy is always lower for the hairy configurations and scales like $(T_c - T)^2$ near $T_c$, showing that 
      this is a second order phase transition. Actually, you can compare the free energy at fixed charge or fixed chemical potential. These correspond to two different ensembles, but in both cases, the free energy is lower for the hairy configuration. The net result is that Reissner-Nordstrom AdS is unstable at low temperatures in the theory (\ref{action}) and develops scalar hair.  

We next want to compute the conductivity as a function of frequency in the dual theory. According to the gauge/gravity duality dictionary, this is obtained by perturbing the Maxwell field around the black hole.  We impose ingoing wave boundary conditions at the horizon since that  corresponds to causal propagation in the dual theory. Assuming harmonic time dependence $e^{-i\omega t}$, we get an ordinary differential equation which can be solved numerically.

Asymptotically,
\be A_x=A_x^{(0)}+{A_x^{(1)}\over r}+\cdots\ee
The gauge/gravity dictionary says the limit of the electric field in the bulk is the electric field on the boundary: $E_x = - \dot A_x^{(0)}$, and the expectation value of the induced current is the first subleading term: $ J_x = A_x^{(1)}$. From Ohm's law we get:
\be\label{eq:conductivity}
\sigma(\w) = \frac{ J_x }{E_x} = - \frac{  J_x }{\dot A_x^{(0)}} = -\frac{ i  J_x }{\w
A_x^{(0)}} = - \frac{i A_x^{(1)}}{\w A_x^{(0)}} \,.
\ee
So the conductivity is directly related to the ratio of the leading two terms at large $r$ of a  perturbation of the Maxwell field. This is typical of how gauge/gravity duality works. The leading term often corresponds to adding a source (in this case, the electric field), and the first subleading term corresponds to the response (the current).\footnote{The perturbation of the Maxwell field induces a perturbation of the metric whose dual interpretation is simply that the current carries nonzero momentum.}

\begin{figure}\begin{center}
\includegraphics[width=.7\textwidth]{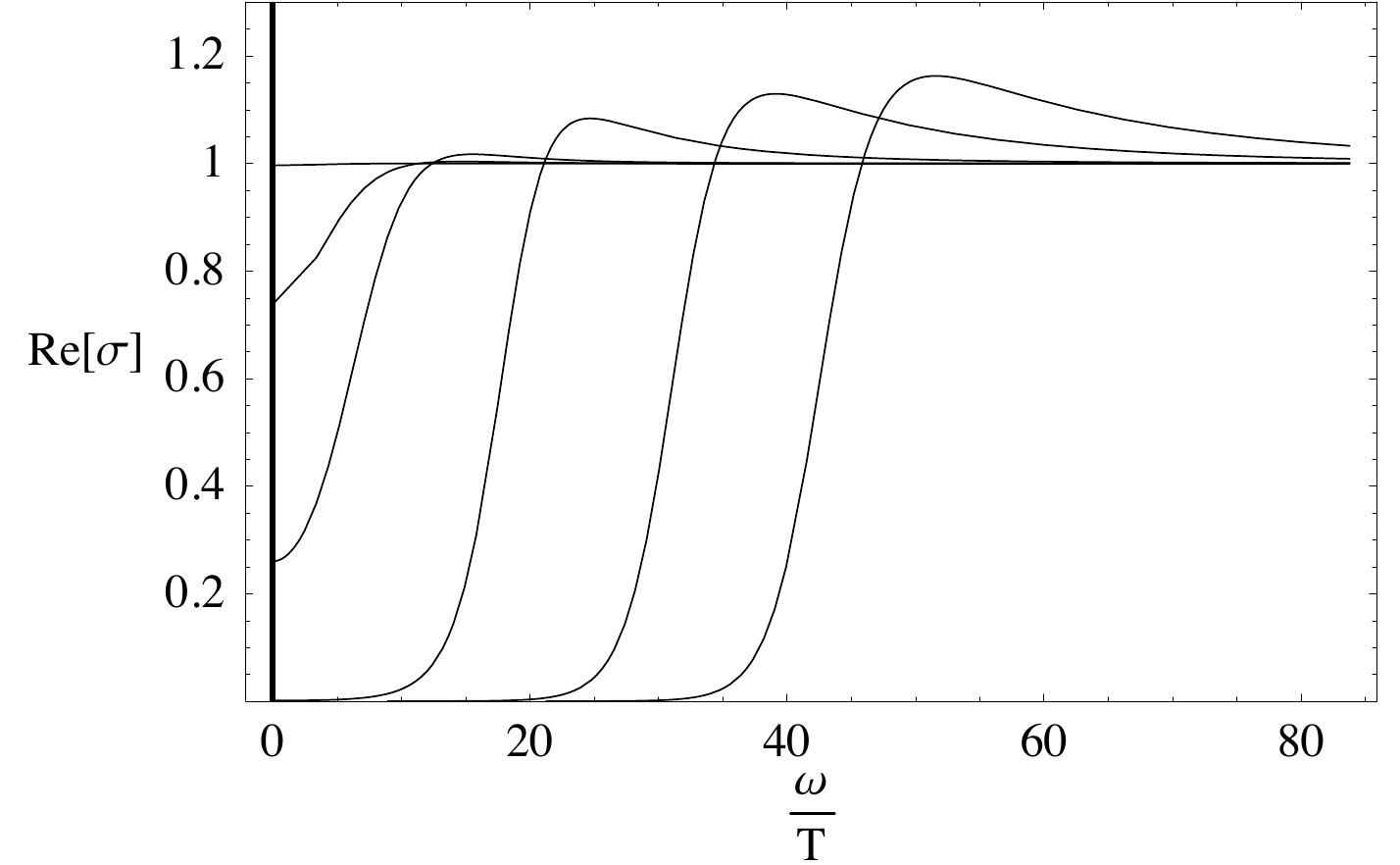}
\caption{The formation of a gap in the real
part of the conductivity as the temperature is lowered below the
critical temperature.  The curves describe successively lower temperatures. There is also a
delta function at $\w = 0$.}\label{fig:gap}
\end{center}\end{figure}

For $T>T_c$, the conductivity is constant. This is a property of theories with gravity duals. As one lowers the temperature below $T_c$, a pronounced gap appears at low frequency (see Fig. 2).\footnote{It was originally thought that the conductivity was behaving like $\sigma \sim e^{-\Delta/T}$ as expected for theories with a mass gap $\Delta$. However, it is now known that the conductivity remains nonzero, although exponentially small, at $T=0$ and low frequency.} There is also a
 delta function at $\w = 0$ for all $T < T_c$ corresponding to infinite DC conductivity. This cannot be seen by numerically solving for the real part of the conductivity, but can be inferred from the behavior of the imaginary part.
 The Kramers-Kronig relations relate the real and imaginary parts of any causal quantity, such as the conductivity, when expressed in frequency space. They imply that the real part has a delta function if and only if the imaginary part has a pole.
One indeed finds a pole in Im$[\sigma(\omega)]$ at $\w = 0$ for all $T < T_c$.

At very low temperatures, one can compare the width of the gap in the optical conductivity $\omega_g$  to the critical temperature.  Even though the model has two free parameters (the mass $m$ and charge $q$ of the scalar field) one finds that this ratio is approximately constant  and given by \cite{Horowitz:2008bn}
\be {\omega_g\over T_c} \approx 8
\ee
 This holds provided $q$ is not too small, in which case the gap becomes much less pronounced and $\omega_g$ is ill defined. (It is also modified by higher curvature modifications to general relativity \cite{Gregory:2009fj,siani}.) 
 In BCS theory, this ratio is about 3.5, but in some of the
high $T_c$ cuprates this ratio has been measured and is indeed about eight \cite{Gomes:2007}. 

This is encouraging, but this simple model certainly does not capture the whole story. One problem is that it has the wrong symmetry. The symmetry of a superconductor refers to the energy gap near the Fermi surface. The fact that our bulk solution is rotationally invariant means that it describes an s-wave superconductor.
High $T_c$ superconductors are known to be d-wave. There has been some recent work trying to construct a d-wave holographic superconductor \cite{Benini:2010pr}, but this remains one of the main open problems in the field.
Interestingly enough, there is evidence that the new iron based superconductors may be s-wave.

There have been many further developments over the past two years. For example,  the response to magnetic fields has been studied.  One of the characteristic properties of superconductors is that they expel magnetic fields. However, a
 superconductor can expel a magnetic field only up to a point. A sufficiently strong  field will destroy the superconductivity. Superconductors are divided into two classes depending on  how they make the transition from a superconducting to a normal state as the magnetic field is increased. In type I superconductors, there is a first order phase transition at a critical field $B_c$, above which magnetic field lines penetrate uniformly and the material no longer superconducts.  In  type II superconductors, there is a more gradual second order phase transition.  The magnetic field starts to penetrate the superconductor in the form of vortices with quantized flux at a lower critical field $B_{c1}$.   The vortices become more dense as the magnetic field is increased, and at an upper critical field strength, $B_{c2} > B_{c1}$, the material ceases to superconduct.
 It was shown that holographic superconductors are type II, just like the high $T_c$ cuprates. A vortex solution has been constructed \cite{Montull:2009fe,Albash:2009iq}.

To summarize, we have seen that  black holes in AdS behave exactly like a  thermal system in one lower dimension.
Gauge/gravity duality predicts this is not a coincidence, but part of a much deeper connection between general relativity and nongravitational physics.
We have seen that one can indeed recover aspects of superconductivity from general relativity.

Carrying this further, one might speculate that someday  one might use gravity to predict new exotic states of matter and then try to look for them in the lab.  One can also use the duality in reverse. Since classical gravity corresponds to a certain (large N) limit of the dual field theory which is hard to realize in condensed matter systems, a connection with quantum gravity might be easier to arrange. It is certainly a striking thought that we
 might someday be able to learn about quantum gravity by doing condensed matter experiments!

\vskip .5cm
\centerline{\bf Acknowledgements}
\vskip .5 cm
I would like to thank the organizers of the GR19 conference for a very stimulating meeting. This work is supported in part by NSF grant PHY-0855415.

\end{document}